\documentclass{emulateapj}
\slugcomment{Accepted by ApJ}
\newcommand{\swift}{{\em Swift}}
\newcommand{\maxi}{{\em MAXI}}
\newcommand{\integral}{{\em INTEGRAL}}
\newcommand{\rxte}{{\em RXTE}}

\shorttitle{Swift Observations of MAXI~J1659$-$152}
\shortauthors{Kennea et al.}

\begin{document}

\title{Swift Observations of MAXI~J1659$-$152: A Compact Binary with a Black Hole Accretor}

\author{J.\,A.~Kennea\altaffilmark{1}, P.~Romano\altaffilmark{2},
  V.~Mangano\altaffilmark{2}, A.\,P.~Beardmore\altaffilmark{3},
  P.\,A.~Evans\altaffilmark{3}, P.\,A.~Curran\altaffilmark{4},
  H.\,A.~Krimm\altaffilmark{5,6}, C.\,B.~Markwardt\altaffilmark{7}, K.\,Yamaoka\altaffilmark{8}}

\altaffiltext{1}{Department of Astronomy \& Astrophysics, The Pennsylvania State
  University, 525 Davey Lab, University Park, PA 16802, USA}
\email{kennea@swift.psu.edu}

\altaffiltext{2}{INAF, Istituto di Astrofisica Spaziale e Fisica Cosmica, Via
  U. La Malfa 153, I-90146 Palermo, Italy}

\altaffiltext{3}{Department of Physics and Astronomy, University of Leicester,
  Leicester, LE1 7RH, UK}

\altaffiltext{4}{AIM, CEA/DSM - CNRS, Irfu/SAP, Centre de Saclay, Bat. 709, FR-91191
Gif-sur-Yvette Cedex, France}

\altaffiltext{5}{Center for Research and Exploration in Space Science and
  Technology (CRESST), NASA Goddard Space Flight Center, Greenbelt, MD
  20771, USA}

\altaffiltext{6}{Universities Space Research Association, 10211 Wincopin Circle, Suite 500,
Columbia, MD 21044-3432}

\altaffiltext{7}{Astroparticle Physics Laboratory, Mail Code 661, NASA Goddard
  Space Flight Center, Greenbelt, MD 20771, USA}

\altaffiltext{8}{Department of Physics \& Mathematics, Aoyama Gakuin University,
  Sagamihara, Kanagawa 252-5258, Japan}

\begin{abstract}

  We report on the detection and follow-up high cadence monitoring
  observations of MAXI~J1659$-$152, a bright Galactic X-ray binary
  transient with a likely black-hole accretor, by \swift\ over a 27\,day
  period after its initial outburst detection.  MAXI~J1659$-$152 was
  discovered almost simultaneously by \swift\ and \maxi\ on 2010 September
  25, and was monitored intensively from the early stages of the outburst
  through the rise to a brightness of $\sim0.5$\,Crab by the \swift\ X-ray,
  UV/Optical, and the hard X-ray Burst Alert Telescopes.  We present
  temporal and spectral analysis of the \swift\ observations. The broadband
  light-curves show variability characteristic of black-hole candidate
  transients.  We present the evolution of thermal and non-thermal
  components of the $0.5-150$ keV combined X-ray spectra during the
  outburst. MAXI~J1659$-$152 displays accretion state changes typically
  associated with black-hole binaries, transitioning from its initial
  detection in the Hard State, to the Steep Power-Law State, followed by a
  slow evolution towards the Thermal State, signified by an increasingly
  dominant thermal component associated with the accretion disk, although
  this state change did not complete before \swift\ observations ended.  We
  observe an anti-correlation between the increasing temperature and
  decreasing radius of the inner edge of the accretion disk, suggesting
  that the inner edge of the accretion disk in-falls towards the black-hole
  as the disk temperature increases. We observed significant evolution in
  the absorption column during the initial rise of the outburst, with the
  absorption almost doubling, suggestive of the presence of an evolving
  wind from the accretion disk. We detect quasi-periodic oscillations that
  evolve with the outburst, as well as irregular shaped dips that recur
  with a period of $2.42\pm0.09$\,hours, strongly suggesting an orbital
  period that would make MAXI~J1659$-$152 the shortest period black-hole
  binary yet known.

\end{abstract}


\section{Introduction}

The majority of stellar mass black-hole candidate binaries (BHB) in our
Galaxy emit mainly in a low level quiescent state. These systems
occasionally go through bright X-ray and optical transient outbursts that
typically last days to months \citep{Chen97}. Although BHB outbursts are
known to be recurrent, the time between outbursts has been reported to be
as long as 60 years \citep{Eachus76}, suggesting many systems remain
undiscovered.  With the use of wide field X-ray monitoring, we are able to
detect and localize these bright transient outbursts, leading to the
discovery of previously undetected BHBs. The X-ray signatures of BHBs are
well known and categorized into distinct states, defined by the presence of
quasi-periodic oscillations (QPOs), the total rms variability of their
power-spectra, and the fitted parameters and relative strengths of thermal
and non-thermal components of their X-ray spectra.  A review of BHB systems
is given by \cite{McandRem06} and for consistency we use the nomenclature
presented in that review to classify emission states.

The combination of \swift\ \citep{Gehrels04}, with its wide field hard
X-ray detector and broadband narrow field instruments, and the regular
almost all-sky X-ray scanning performed by {\em ``Monitor of All-sky
  X-ray Image''} (\maxi; \citealt{Matsuoka09}), allows for rapid
detection and accurate localization of BHBs as they enter
outburst. \swift's low-overhead observing and rapid autonomous
follow-up capabilities allow us to obtain high cadence monitoring
observations in X-ray and Optical/UV wavelengths quickly after initial
detection, allowing the study of the earliest stages of BHB outbursts
with new clarity.

MAXI~J1659$-$152 was first reported after detection by the \swift\ Burst Alert
Telescope (BAT; \citealt{Barthelmy05}) at 08:05 UT, 2010 September 25
(MJD 55464.337, all times from this point are quoted using MJD format
in UTC). Follow up observations performed by the \swift\ X-ray
Telescope (XRT; \citealt{Burrows05}) and UV/Optical Telescope (UVOT;
\citealt{Roming05}) 31 minutes later localized the transient
\citep{Mangano10}, although it was initially misidentified as a
Gamma-Ray Burst and named GRB 100925A.  Based on its detection by the
\maxi\ Nova Alert System \citep{Negoro09} at MJD 55464.104 (02:30UT,
$\sim5.5$\,hours before the BAT trigger), MAXI~J1659$-$152 was determined to be
a previously unknown Galactic X-ray transient \citep{Negoro10}.

IR spectroscopy was obtained which confirmed that the optical counterpart
showed emission lines consistent with that of an X-ray Binary
\citep{deUP10}. The transient was also detected in radio \citep{vdH10}, by
\integral\ \citep{vovk10}, {\em XMM-Newton}\ \citep{Kuulkers10a} and \rxte,
which detected a 1.6 Hz type-C QPO in the power-spectrum, indicating that
MAXI~J1659$-$152 is a BHB \citep{Kalamkar11}.  \cite{Kuulkers10b} reported
evidence for periodicity in the 2.4--2.5\,hour range from {\em XMM-Newton}\
data, suggesting that this is the shortest period BHB yet
known. \cite{Belloni10} reported a refined the period measurement from
\rxte\ data of $2.4142$\,hours. The {\em XMM-Newton} light-curve revealed
irregular structure dips lasting $5-40$\,min, and suggest that dips
analogous to those often seen in Low Mass X-ray Binaries (LMXBs) are the
source of the $\sim2.4$\, hour period, rather than eclipses from the
companion star \citep{Kuulkers10b}.

We report here on broadband observations of MAXI~J1659$-$152 utilizing all
three instruments on \swift\ during the first 27\,days of the outburst
after its initial detection. We present spectral and temporal
analysis, including the broadband UV/optical, X-ray and hard X-ray
light-curves, analysis of QPOs, time resolved spectral evolution
utilizing broadband spectral fits across the XRT and BAT energy
ranges, and search for periodicities in the X-ray data.

\section{Observations and Analysis}

Observations with \swift\ began after the rising hard X-ray brightness of
MAXI~J1659$-$152 triggered the BAT at MJD 55464.337. This prompted the
standard GRB follow-up mode (e.g.  \citealt{Gehrels04}) in which the
transient was observed as an ``Automated Target'' (AT) every
$\approx96$\,minute orbit with exposures of 0.5--2.5\,ks per orbit.  From
MJD 55468 onwards the observation cadence was lowered two 1--2\,ks
observations a day, approximately spaced by 12\,hours. Observations of the
source continued with this cadence until the final observation ended at MJD
55491.259, $\sim27$\,days after the initial BAT detection, after which
MAXI~J1659$-$152 became too close to the Sun for \swift\ to
observe. Between MJD 55480 and MJD 55482 MAXI~J1659$-$152 was not
observable by XRT or UVOT due to the proximity of the Moon. XRT observed in
Windowed Timing (WT) mode for all observations except for a 1\,ks observation
taken on MJD 55466 in Photon Counting (PC) mode, which was performed to
obtain an accurate localization. UVOT data were typically collected
utilizing all 6 UVOT filters, apart from a period between MJD 55468 and MJD
55479 when observations were taken utilizing a daily rotation of the 3 UV
filters and $u$.  MAXI~J1659$-$152 was observed by \swift\ for 1\,ks on MJD
55598, 107 days after the initial monitoring observations ended, with XRT
data collected in PC mode, and UVOT utilizing all 6 filters.

BAT data from the observations during which \swift\ was pointed at
MAXI~J1659$-$152 were processed using the HEASOFT {\ttfamily batsurvey}
script to produce eight-channel light curves which were then converted to
spectra covering the energy range $14-195$\,keV. BAT light-curves were
produced automatically by the BAT Transient Monitor web page
\citep{Krimm06}.

XRT light-curves and spectra were extracted utilizing the methods
described by \cite{Evans09}, with full corrections for pile-up and hot
columns applied to the data based on a PSF fitted position of MAXI~J1659$-$152
obtained from PC mode observations.  XRT spectra were extracted over 
time intervals strictly simultaneous with the BAT survey spectra, and 
binned to a minimum of 20 counts per energy bin.

UVOT photometry was derived from images via {\tt uvotmaghist}, using an
extraction region of radius 5\arcsec.  Magnitudes are based on the UVOT
photometric system \citep{poole2008:MNRAS383} and were uncorrected for the
Galactic extinction in the direction of MAXI~J1659$-$152 of $E_{(B-V)} =
0.606$ \citep{schlegel1998:ApJ500}.  Taking this value as an upper limit to
the extinction of the counterpart and using the effective wavelengths of
the \swift\ filters \citep{poole2008:MNRAS383} and the
parameterization of \cite{Pei92}, the extinctions in the
\swift\ bands are: $A_{v} \leq 1.85$, $A_{b} \leq 2.36$, $A_{u} \leq
2.96$, $A_{uvw1} \leq 4.10$, $A_{uvw2} \leq 4.88$, $A_{uvm2} \leq 5.84$.
Magnitudes of the UVOT photometric system are indicated by lower-case
letters of the filter used (e.g. $v$), and catalog magnitudes by upper-case
letters of the filter (e.g. $V$). However within UVOT measurement errors,
we can assume that UVOT $u$, $b$ and $v$ magnitudes are equivalent to $U$,
$B$ and $V$.

All quoted uncertainties are given at $90\%$ confidence level for one
interesting parameter unless otherwise stated.

\section {Results}
\subsection {Localization}

Utilizing a short XRT observation in PC mode taken on MJD 55598, we derived
a position of RA, Dec = $16^h59^m01^s.71$, $-15^\circ15'28''.5$ (J2000,
1.4\arcsec\ error radius).
This position was corrected for systematic errors in astrometry utilizing
UVOT data taken simultaneously with the XRT data, utilizing the methods
described by \cite{Goad07}. This position is improved over the
previously reported XRT position \citep{Kennea10}, which was based on PC mode
data taken on MJD 55466 when MAXI~J1659$-$152 was bright and the data were highly
affected by pile-up. During the later observation MAXI~J1659$-$152 was much fainter and the
data where not affected by pile-up, thus allowing for a more accurate
localization.

The UVOT position of the counterpart was derived from a deep (7170\,s) $v$
band image summed between MJD 55464.6 and 55467.6, using the {\tt
  uvotdetect} command, as RA, Dec = $16^h59^m01^s.679$,
$15^\circ15'28''.54$ (J2000; 0.70\arcsec\ error radius), 0.4\arcsec\ from
the center XRT position and within the XRT error circle, and 0.19\arcsec\
from the EVLA detected radio counterpart \citep{Paragi10}, unambiguously
confirming that the UVOT source is the optical counterpart of the
MAXI~J1659$-$152.
The UVOT location of MAXI~J1659$-$152 does not correspond with any known catalog
object. The non-detection of the companion in the USNO-B catalog allows us
to place an upper limit on the optical brightness of MAXI~J1659$-$152 in quiescence
of $V>21$ \citep{Monet03}.

\subsection{Broadband Outburst Light-curve}

\begin{figure}
\resizebox{\hsize}{!}{\includegraphics[angle=0]{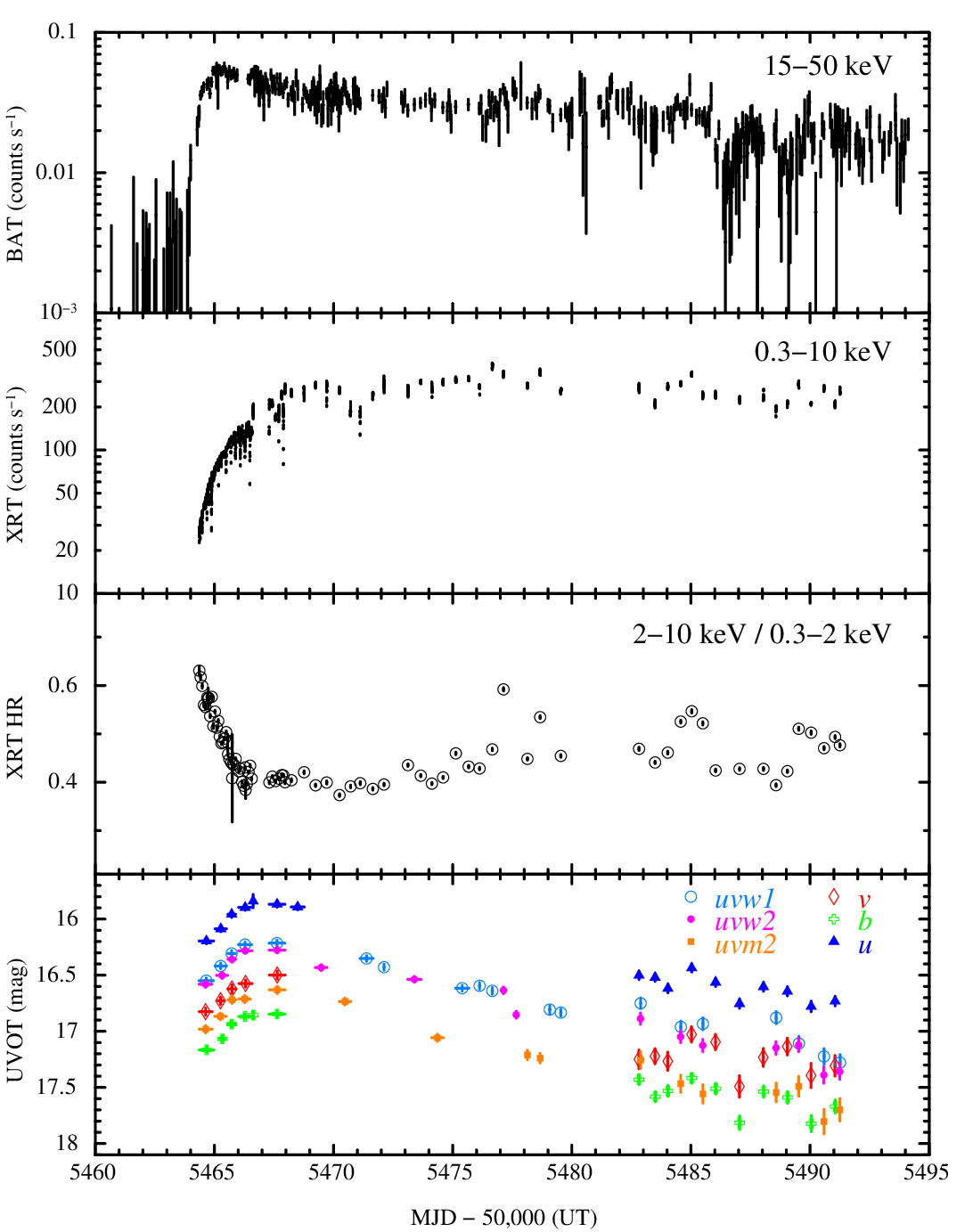} }
\caption{Outburst light-curves of MAXI~J1659$-$152. From top to bottom: $15-50$\,keV BAT
  Transient Monitor light-curve with orbital binning, filtering out
  bins with less than 500s integration times; $0.3-10$\,keV XRT count rate 
  light-curve with 100s time bins; Ratio of $2-10$\,keV and $0.3-2$\,keV XRT count rates,
  binned by orbit; UVOT six filter light-curve binned by observing segment. Errors are $1\sigma.$}
\label{fig:light-curves}
\end{figure}

Figure~\ref{fig:light-curves} shows the XRT, BAT and UVOT light-curves for
MAXI~J1659$-$152 during the initial stages of the outburst.  Examination of
data from the \swift/BAT hard X-ray transient monitor show that the
outburst light-curve has the typical fast-rise/exponential decay (FRED)
shape seen in many X-ray transients. The source had risen very sharply from
non-detection on MJD 55462 to a weak detection on MJD 55463 ($0.0029 \pm
0.0012$ counts\,cm$^{-2}$\,s$^{-1}$, $15-50$\,keV) and then a $26\sigma$\
detection ($0.033 \pm 0.0012$ counts\,cm$^{-2}$\,s$^{-1}$) in the daily
average on MJD 55464. There was no detection of the source dating back to
January 2006 with a $3\sigma$\ upper limit of $0.0045$ counts\,cm$^{-2}$
s$^{-1}$. The BAT light-curve is well modeled by a sharp rise from a low
level detection to peak rate of $0.0521 \pm 0.0019$
counts\,cm$^{-2}$\,s$^{-1}$ on MJD 55465, followed by a slow exponential
decay, until it became undetectable in the BAT $15-50$\,keV energy range
after MJD 55529, with a $3\sigma$ upper limit of $0.008$\,counts\,s$^-1$\,
cm$^2$.

The XRT WT light-curve starts on MJD 55464, at a level of $\sim26$
counts\,s$^{-1}$, which corresponds to a flux of $1.9 \times 10^{-9}$ erg
s$^{-1}$\,cm$^{-2}$ ($0.5-10$\,keV).  The light-curve continues to rise to
peak around MJD 55476, approximately 11\,days after the peak of the BAT
light-curve, after which MAXI~J1659$-$152 begins a slow decline.  We have
plotted the ratio of two hard and soft XRT bands, $2-10$\,keV and
$0.3-2$\,keV, respectively. This hardness ratio plot shows that for the
initial phase of the outburst of MAXI~J1659$-$152 there is a rapid spectral
softening which reaches a soft extremum on MJD 55657, after which the
hardness shows a slow increase, with some short term variability. The
abrupt change in the hardness ratio on MJD 55657 is most likely a
indication of the evolution of the source from the Hard State, to the Steep
Power-Law state, which is followed by a slow evolution into the Thermal
State.

The UVOT counterpart shows significant variability over all ($v$, $b$, $u$,
$uvw1$, $uvw2$, $uvm2$) bands. The UVOT light-curve shows a correlated rise
with the X-ray light-curve, peaking around MJD 55465 at $v=16.5$, followed
by a slower decline.  

The observation performed on MJD 55598 (not shown in
Figure~\ref{fig:light-curves}) finds the source at a greatly lowered flux
level than seen during outburst, approximately $0.52 \pm 0.02$ counts\,s$^{-1}$, 
corresponding to a flux of $2.4 \pm 0.1 \times 10^{-9}\ \mathrm{erg\ cm^{-2}\ s^{-1}}$. The observed hardness ratio ($0.74 \pm 0.04$) during
this observation suggests that the source had returned to the Hard
State. The UVOT data show that the optical counterpart had faded to $v=18.6
\pm 0.3$ mag, a drop of approximately 1.3 magnitude from the previous
measurement 107 days earlier, and $2.1$ magnitudes fainter than the peak
outburst brightness. As discussed earlier, we assume that the quiescent 
optical brightness is $V>21$, and therefore on MJD 55598 MAXI~J1659$-$152 is still
in outburst.

\subsection{Spectral evolution}

\begin{figure}
\resizebox{\hsize}{!}{\includegraphics[angle=270]{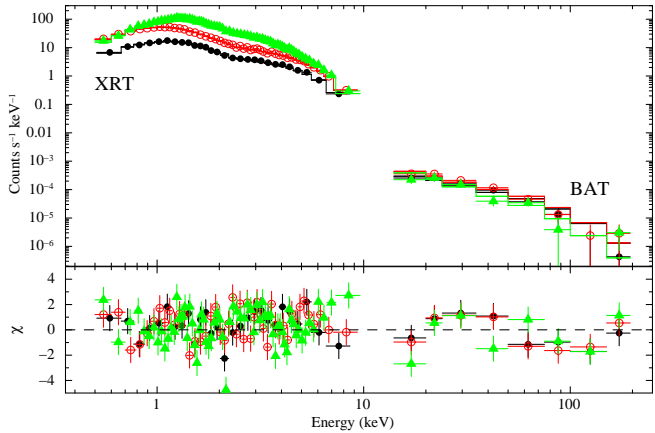}}
\caption{Combined BAT$+$XRT spectra of MAXI~J1659$-$152 with integrations
  centered on three different epochs during its outburst: MJD 55464.42
  (black dots), MJD 55465.23 (red circles) and MJD 55484.57 (green triangles). }
\label{fig:3spec}
\end{figure}

We fit the simultaneous BAT$+$XRT spectra in the $0.5-10$\,keV and
$15-150$\,keV energy bands for XRT and BAT, respectively.  Factors were
included in the fitting to allow for normalization uncertainties between
the two instruments, which are constrained within their usual ranges
(0.9--1.1).  Each spectrum was fit with a three-component model consisting
of a power-law model ({\tt powerlaw}), a model of thermal emission from the
accretion disk ({\tt diskbb}, \citealt{Makishima86}), and a model of the
total absorption ({\tt tbabs}) with the abundance set to the values given
by \cite{Wilms00}.  We calculated the fraction of flux contribution from
the {\tt diskbb} model, $f^b$, by finding the ratio of the model fitted
$2-10$\,keV flux from the {\tt diskbb} component with the combined flux
from both the {\tt diskbb} and {\tt powerlaw} components. Example XRT$+$BAT
spectra at three different epochs are shown in Figure~\ref{fig:3spec}.

In order to test the requirement of the disk component over a simple
absorbed power-law model, we performed an F-test comparison between fits of
the two models for every fitted spectra throughout the outburst.
For fits before MJD 55465, which coincides with the peak of the BAT
flux (Figure~\ref{fig:light-curves}), we find an F-test probability of
$>10^{-3}$ for approximately $80\%$ of the fitted spectra,
suggesting that during this period, a disk component is not needed with a
high significance to fit the X-ray spectrum, as expected for a BHB in the
Hard State. After MJD 55465, we find that $96\%$ of the fitted spectra have
an F-test probability of $<10^{-9}$, showing that the addition of a disk
component significantly improves the fit for the majority of cases. We
therefore conclude that the disk component is statistically significant for
all times during the \swift\ observations of the outburst other than during
the very early part (earlier than MJD 55465).

\begin{figure}
\resizebox{\hsize}{!}{\includegraphics[angle=0]{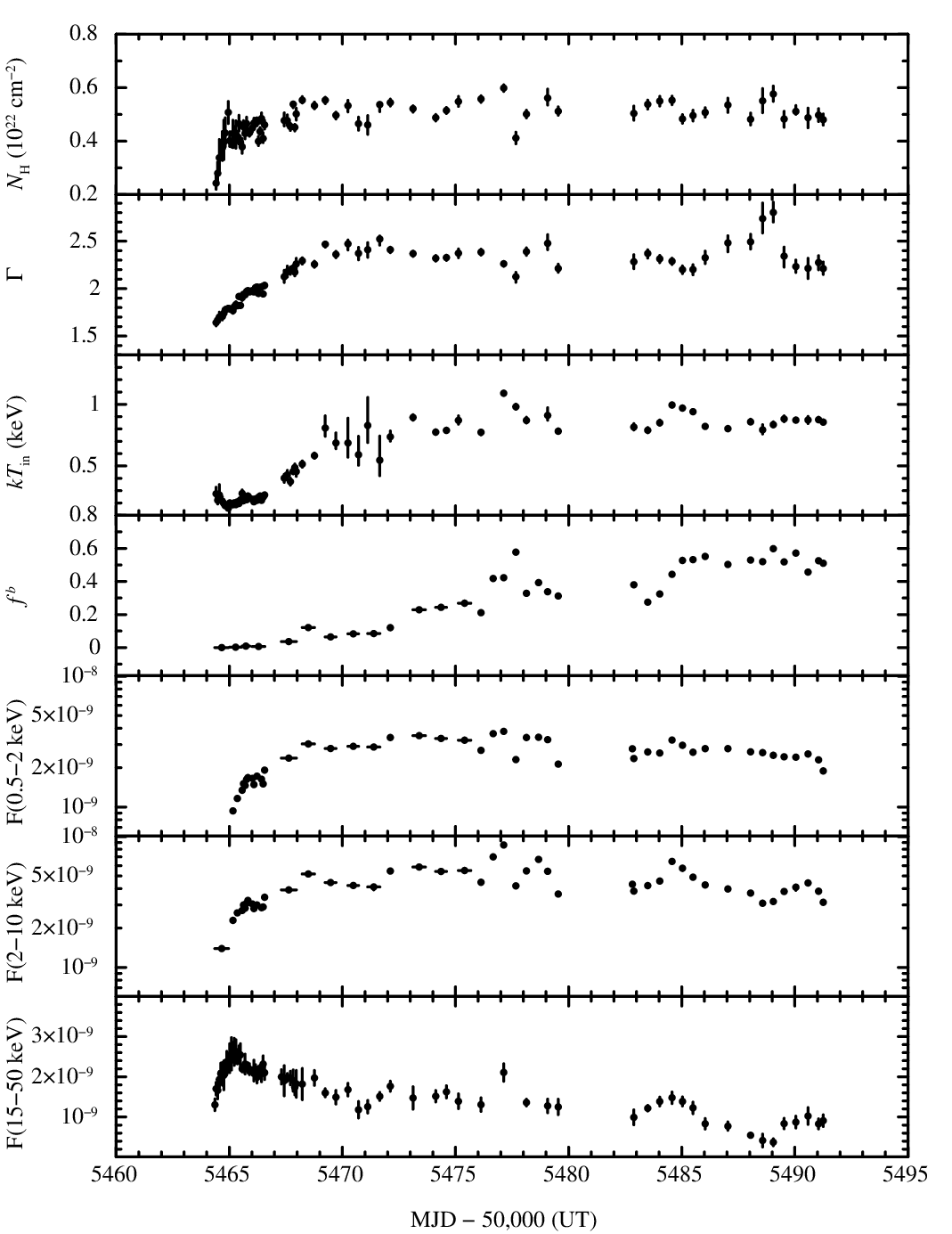} }
\caption{Results of combined BAT$+$XRT spectral fitting of MAXI~J1659$-$152 with
  an absorbed power-law plus disk-blackbody component. Derived spectral
  parameters are (from top to bottom), Hydrogen column density
  ($N_\mathrm{H}$), power-law photon index ($\Gamma$), disk-blackbody
  inner disk temperature ($kT_\mathrm{in}$), disk fraction ($f^b$, the
  fraction of $2-10$\,keV flux contributed by the disk component),
  and the $0.5-2$\,keV, $2-10$\,keV and $15-50$\,keV fluxes.}
\label{fig:spectral_evo}
\end{figure}

The fitted spectral parameters as a function of time are shown in
Figure~\ref{fig:spectral_evo}. As inferred from the hardness ratio changes,
MAXI~J1659$-$152 shows significant spectral evolution during the
outburst. During the initial detection the spectrum is well fit by an
absorbed power-law model with $\Gamma \simeq 1.8$, indicating a BHB in the
Hard State.  As MAXI~J1659$-$152 brightens the spectrum softens and
continues to be dominated by power-law emission with a $\Gamma \simeq 2.4$
on MJD 55468, signaling the transition to the Steep Power-Law state. During
this time, the disk blackbody component is a small ($f^b < 10$\%), but
statistically significant and evolving ($kT_\mathrm{in}$ rising from
$\sim0.2$\,keV to $\sim0.6$\,keV), component of the spectrum.

After MJD 55469 the source appears to evolving to the Thermal State,
indicated by the disk fraction $f^b$ increasing from $\sim0.1$ on MJD
55465 to $\sim0.5$ by MJD 55485 and remaining approximately constant until
the proximity of MAXI~J1659$-$152 to the Sun forced the end of \swift\ monitoring
observations on MJD 55491. In this period $kT_{\mathrm{in}}$ is in the
range of $0.8-1$\,keV, typical for BHB in the Thermal State. Therefore we
do not observe the full state transition between the Steep Power-Law and
Thermal states, defined by \cite{McandRem06} as occurring when $f^b >
75\%$, with the source remaining in an intermediate state.

In order to assess the possible presence of a spectral break, we fit a
model to the data consisting of a {\tt diskbb} plus cut-off power-law ({\tt
  cutoffpl}) model to data from MJD 55484.57 (see Figure~\ref{fig:3spec})
chosen as this data correspond to the peak BAT flux of
MAXI~J1659$-$152. The fitted value of the cut-off energy is poorly
constrained, $E_\mathrm{cut}=88.20_{-42.89}^{+111.81}$ keV, with a measured
photon index of $\Gamma=2.14_{-0.16}^{+0.15}$. An F-test shows that there
is little evidence of an improvement in the fit if a spectral break is
added (F=2.02, giving a null-hypothesis probability of 0.16), and therefore
the cut-off power-law is not required to fit the XRT$+$BAT data. We note
that there are currently no analysis tools available that allow integration
of BAT spectra over longer periods than an individual \swift\ observation
snapshot, which would be required to improve the accuracy of the spectral
measurement. Given this and the brightness of the BAT data, if a spectral
break is present, the XRT$+$BAT data are not sensitive to it. 

The expected Galactic line-of-sight absorption for MAXI~J1659$-$152 is $1.7
\times 10^{21}$\,cm$^{-2}$ \citep{Kalberla05}, however the fitted
absorption column is higher and variable, indicating an additional
intrinsic absorption component. The fitted $N_\mathrm{H}$ increases rapidly
during the initial stages of the outburst, beginning at $N_\mathrm{H} = 2.4
\pm 0.3 \times 10^{21}\ \mathrm{cm^{21}}$ on MJD 55464 and rising to a mean
value of $N_\mathrm{H} = 5 \times 10^{21}\ \mathrm{cm^{2}}$ on MJD
55465. Measured absorption remains consistent with this value until
observations ceased on MJD 55491.  Fitting more complex models for
absorption such as a warm absorber or partial covering model, with a fixed
Galactic absorption also provide good fits to the increasing absorption,
although given the spectral resolution of the XRT data, these more complex
models are not required over a simple single-parameter variable absorption
model.

A later follow-up observation taken on MJD 55598 after MAXI~J1659$-$152 had faded
significantly to a flux of $2.4 \pm 0.1 \times 10^{-11}\ \mathrm{erg\
  s^{-1}\ cm^{-2}}$ (0.5-10 keV) and entered the Hard State ($\Gamma =
1.81\pm0.1$, no significantly detected disk component), gives a fitted
absorption of $N_\mathrm{H} = 4.2 \pm 0.6 \times 10^{21}\
\mathrm{cm^{-2}}$, consistent with the post-outburst value, suggesting that
the additional absorption component is still present despite the lowered
X-ray flux.

With moderate resolution spectral fitting, variable absorption may be due
to a statistical correlation between spectral parameters. To investigate
this we examined confidence contours in ($N_\mathrm{H}$, $\Gamma$)
parameter space, and find that the low $N_\mathrm{H}$ spectra (MJD 55464)
are distinct from the higher $N_\mathrm{H}$ spectra (MJD 55465) with
$5\sigma$ confidence.  Joint BAT and XRT spectral fits allow to constrain
$\Gamma$ and $N_\mathrm{H}$ more independently than by fitting XRT alone.
Care was taken to remove the effects of pile-up from the spectra, as
pile-up can considerably affect continuum spectra fitting \citep{Miller10}.
The fitted values of $N_\mathrm{H}$ and $\Gamma$ do not vary in lock-step
either, given the fast initial rise in $N_\mathrm{H}$ between MJD 55464 and
MJD 55465, compared with an linear rise in $\Gamma$ ending around MJD
55470.  The consistency of the measured absorption in the later follow-up
observation of MAXI~J1659$-$152 provides further evidence that the measured
$N_\mathrm{H}$ is not related to the fitted model or source brightness.  We
therefore conclude that the variable absorption is physical and not an
artifact of the fit.

\subsection{Time series analysis}

We searched for QPOs using a fast Fourier transform method, adopting
the power-spectral normalization of \cite{Leahy83} and the search
technique described by \cite{vanderKlis89}. For each orbit of XRT
data, power density spectra (PDS) were generated for $M$ continuous
sections of data of 4096 bins duration, with a bin size of 0.01 s, and
averaged.  The averaged PDS from each orbit was then rebinned in
frequency so that $W$ continuous frequencies were averaged, using a
geometrical series binning scheme.

\begin{figure}
\resizebox{\hsize}{!}{\includegraphics[angle=270]{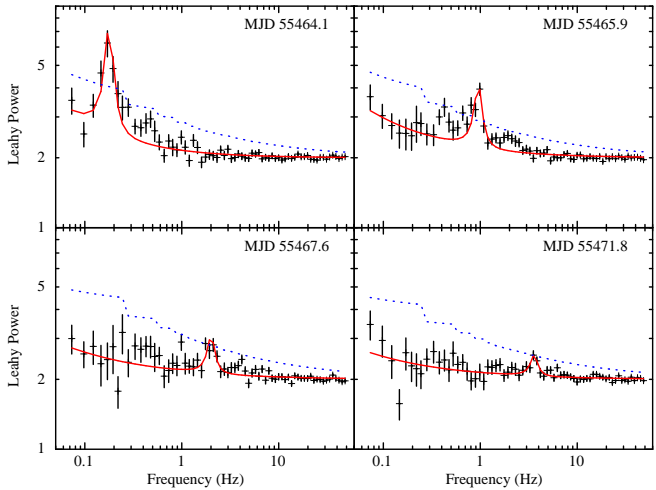} }
 \caption{Example PDS for 4 epochs of the MAXI~J1659$-$152 outburst. The red
   line shows the fitted continuum + Lorentzian model, the blue
   dotted line the $5\sigma$ detection limit for the QPO.}
\label{fig:4_pspe}
\end{figure}

Each rebinned/averaged PDS was fit with a model consisting of a
power-law for the low frequency noise, a Lorentzian for the QPO (whose
width was fixed to preserve a quality factor $Q = \nu / \Delta\nu$ of
5) and a constant for the Poisson noise level. The QPO was considered
detected if the Lorentzian amplitude exceeded the local detection
level, $P_\mathrm{detect}(MW)$ (given by the integral probability of a
chi-squared distribution for $2MW$ d.o.f,. scaled by a factor of 1/MW)
minus the mean noise level, which is 2 for the \cite{Leahy83}
normalization. Example PDS for 4 epochs are shown in
Figure~\ref{fig:4_pspe}.

\begin{figure}
\epsscale{.80}
\resizebox{\hsize}{!}{\includegraphics[angle=270]{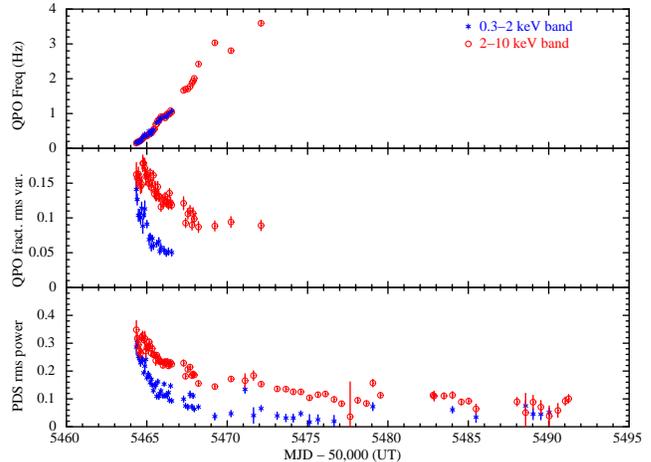} }
 \caption{Results of PDS analysis on XRT WT data two energy bands,
   $0.3-2$\,keV and $2-10$\,keV. For all 5$\sigma$ detections of a
   QPO, we plot the QPO frequency (top), the fractional rms variability of
   the QPO (middle) and the total rms variability in the $0.02 -
   10$\,Hz band including the QPO (bottom). }
\label{fig:qpo_freq}
\end{figure}

Figure~\ref{fig:qpo_freq} shows in two energy bands ($0.3-2$\,keV and
$2-10$\,keV), the QPO frequency, the fractional rms variability of the
fitted Lorentzian for PDS in which QPO were detected at greater than
$5\sigma$, and the broadband continuum rms for $0.02-10$\,Hz, including the
QPO and low-frequency noise components.  

The QPO frequency and detection energy evolves with the QPO peak frequency
increasing approximately linearly with time. After MJD 55466 QPOs are not
detected in the $0.3-2$\,keV band, but are still detected in the
$2-10$\,keV band until MJD 55472. The QPO frequency is correlated with
the fitted value of $\Gamma$, which has been observed previously in BHBs
(e.g. 4U~1543-47, \citealt{Kalemci05}).

The PDS rms power for both energy bands fall rapidly during the initial
outburst, as expected given the transition between the Hard State and the
Steep Power-Law State which occurs on MJD 55467. In the Thermal State, QPOs
are generally weak or absent \citep{McandRem06}, therefore the
non-detection of QPOs after MJD 55474 is consistent with the evolution
towards that state.

The XRT detected QPOs are consistent with those seen by \rxte\
\citep{Kalamkar11}. Due to being on-source earlier than than \rxte, we
detect lower frequency QPOs in the range of $0.148 \pm 0.006$ Hz to $1.931 \pm
0.030$\,Hz in the time period of MJD 55464.35 to MJD 55467.88. The
evolution in frequency over time of these QPOs is consistent with the
extrapolation of the trend seen during \rxte\ observations between
MJD 55467 and approximately MJD 55478.

UVOT data were also obtained in event mode between MJD 55472.1--55475.4 in
$uvw1$, $uvw2$, and $uvm2$ bands.  Unbinned source light curves (not
background subtracted), for each of these, were used to produce PDS and to
calculate the rms variability with the HEASOFT tools {\tt powspec} and {\tt
  lcstats} respectively.  Within each of these the binning was set to be
$N$ times the minimum bin size (11.0322\,ms) where $N=1$ for the PDS and
$N=8$ (88.2576\,ms) for the rms variability, giving frequency ranges of
0.006--90\,Hz and 0.006--1\,Hz respectively. Both the PDS and the rms imply
a level of variability consistent with zero, though can only place a loose
constraint of $\lesssim 50\%$.

\begin{figure}
\resizebox{\hsize}{!}{\includegraphics[angle=270]{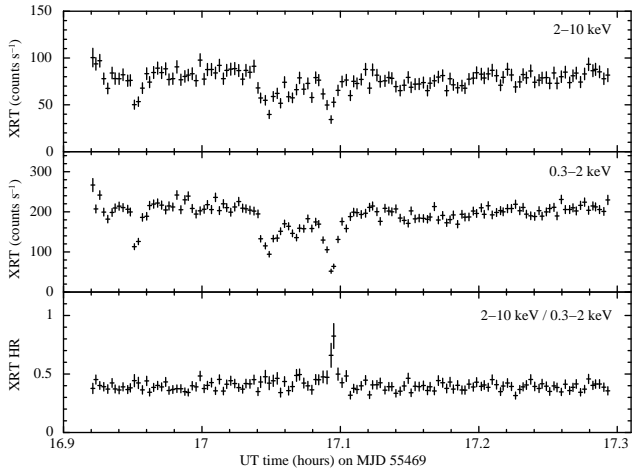} }
 \caption{Example of the hard and non-hard dips in XRT WT mode
   data in two energy bands ($0.3-2$\,keV and $2-10$\,keV), and the
   hardness ratio.}
\label{fig:dip}
\end{figure}

The XRT light-curve shows frequent dips, which do not appear to have a
consistent shape or spectral signature.  An example of a dipping episode, seen
on MJD 55469, is shown in Figure~\ref{fig:dip}, although it should be noted
that this is not necessarily representative of other dips. The dipping
shows two components that we differentiate by their spectral signature: a
series of dips of varying timescales, with no apparent associated spectral
change (non-hard dips), and a narrower sharp dip with an associated
spectral hardening (hard-dip). Analysis of UVOT event data taken during
dips do not show any evidence of correlated optical variability, although
no event mode data was taken during a hard dip.

If, as \cite{Kuulkers10b} suggest, the dips are analogous to those seen in
a LMXB dipping source, we expect the spectral changes during dips can be
explained by the addition of an absorbing component in the spectrum
(e.g. \citealt{DiazTrigo06}). To check this we fit the hard dip spectrum
for the event shown in Figure~\ref{fig:dip} utilizing the same spectral
parameters for the power-law, disk and absorption components for MJD 55469
as shown in Figure~\ref{fig:spectral_evo}, with the addition of a partial
covering absorber (XSPEC {\tt pcfabs} model). We find that the hard dip
spectrum is well fit ($\chi^2 = 39.92$ for $34$ degrees of freedom) by the
addition of an additional partial covering absorber with $N_\mathrm{H} =
3.6^{+2.7}_{-1.6} \times 10^{22}$\,cm$^{-2}$ and a covering fraction of
$0.56 \pm 0.13$. We note that replacing the partial covering with a warm
absorber model (XSPEC {\tt absori} model) provides a slightly worse fit to
the data ($\chi^2 = 40.72$ for 34 degrees of freedom), suggesting that
higher spectral resolution data would be required to determine the nature
of the absorbing material. We therefore conclude that the spectral hardness
changes seen during some of the dips events are likely caused by localized
absorbing material in the disk.

To search for longer timescale periodicities, we used the Lomb-Scargle
method \citep{Lomb76,Scargle82}, which is effective when analyzing unevenly
sampled data. Longer term light-curve trends were removed by fitting a 3rd
order polynomial.  The resultant periodogram, derived from XRT WT data in
the $0.3-10$\,keV band, shows a strong peak at $2.42\pm0.09$\,hours.  This
period is $\sim1.5$ times the $\sim96$ minute \swift\ orbital period,
however given the closeness of our result to the \rxte\ measured
periodicity \citep{Belloni10}, we conclude that it is not an aliasing effect.
No periodicity was detected in the UVOT data from this source, although as
UVOT data were taken in non-event mode for the majority of the
observations, the UVOT data are not sensitive to a 2.4 hr period.

Although periodic, the dip shapes are highly variable, suggesting that
they may be caused by obscuration by optically thick clumps in the
accretion flow, or an evolving bulge in the disk, rather than
eclipsing by the companion star.  Given this scenario, the 2.4\,hour
periodicity is likely the orbital period of MAXI~J1659$-$152.

\section{Discussion}

Without a measurement of $v \sin i$ for MAXI~J1659$-$152 it is difficult to
place strong limits on the orbital parameters of the system from the
\swift\ data alone. However, with knowledge of the period and other system
parameters we can attempt to estimate distance and system geometry from
theoretical and empirically derived relationships that have been
established for LMXBs.

Given that in the X-ray light-curve we see dips but not eclipses, from
the standard model of X-ray binary geometry of \cite{Frank87}, MAXI~J1659$-$152 is
analogous to a ``Pure Dipper'', which constrains the inclination to $i =
60 - 75^\circ$.

\cite{Frank02} derive a simple relationship for the mass of the secondary
star and the orbital period, $M_2\simeq 0.11 P_\mathrm{orb}\mathrm{(hr)}$,
assuming that the companion star is filling its Roche Lobe and is an
unevolved low-mass star. This gives us an estimate of the companion mass of
$M_2 = 0.27\ \mathrm{M_\odot}$. \cite{Smith98} also derive an empirical
linear mass-period relationship, which for $P=2.42 \pm 0.09$\,hours, gives
$M_2 = 0.19 \pm 0.05 \mathrm{M_\odot}$, close to the previously derived
value. Assuming a black-hole primary, $M_1 > 3.2\,\mathrm{M_\odot}$, we estimate an
upper limit on the mass ratio of $q < 0.06$.

The faintest optical magnitude of MAXI~J1659$-$152 seen by UVOT is $v=18.5$ mag on MJD
55598. As the out-of-outburst optical magnitude is fainter than $V>21$,
based on the non-detection in USNO-B catalog, this suggests that the
optical counterpart has not been observed by UVOT in quiescence. Therefore
we can make no estimate, from the UVOT data alone, of the brightness of the
secondary star in this system.  However, if we assume a main sequence
companion, the relationship of spectral type versus period of \citet{Smith98}
allows us to estimate that the companion is a M5 star.

M5 stars have typical absolute visual magnitudes of $M_V \simeq
12-15$. \cite{Patterson84} derive a relationship for absolute visual
magnitude of the companion star for a cataclysmic variable system as a
function of the period, $M_V = 17.7 - 11 \log P_\mathrm{orb}$, assuming this
relationship is valid for a BHB system, we estimate the brightness of the
companion star to be $M_V \simeq 13.5$. Given that $V>21$ in quiescence,
this places a lower-limit on the distance of $d>316$\,pc. Note that this
distance does not include and any extinction correction, which would make
this distance estimate lower.

However, the brightness of the X-ray outburst suggests that
MAXI~J1659$-$152 is much more distant than this, as the peak outburst flux
of $10^{-8}\ \mathrm{erg\ s^{-1}\ cm^{-2}}$ relates to luminosity of
$<0.001\ L_\mathrm{Edd}$ at 631\,pc for a black-hole mass of $M_1 > 3.2\
\mathrm{M_\odot}$, and it is unlikely that the peak outburst would be so
highly sub-Eddington. If we assume the peak flux of MAXI~J1659$-$152
equates to a BHB emitting at $L_{\mathrm X} > 0.1\ L_\mathrm{Edd}$, we
obtain a lower-limit on the distance of $d > 6.1$\,kpc.

\cite{vanParadijs94} derive a relationship between the absolute visual
magnitude of a LMXB as a function of the orbital period and the X-ray
luminosity as a fraction of the Eddington Luminosity, $M_V = 1.57\pm0.24 -
2.27\pm0.32\log\Sigma$, where $\Sigma = (L_\mathrm{X}/L_\mathrm{Edd})^{1/2}
(P_\mathrm{orb}/1hr)^{2/3}$.  If we assume at maximum X-ray luminosity,
$L_\mathrm{X}/L_\mathrm{Edd} = 0.1$, we calculate the maximum absolute
visual magnitude based on this relationship as $M_V = 2.13\pm0.25$. The
peak observed optical brightness for MAXI~J1659$-$152 is $v=16.5$, which
gives estimated distances in the range of $3.2\pm0.5$\,kpc (assuming
Galactic extinction, $A_V = 1.85$) to $7.5\pm0.9$\,kpc (no extinction).

As an alternative measure of distance, we use the \cite{Shahbaz98}
distance-period relationship for X-ray transients. With a peak outburst
optical magnitude of $v=16.5$, an estimated line-of-sight extinction of
$A_v = 1.85$, and assuming the fraction of the optical emission in
quiescence that comes from the companion star, $f$, is $\le 1$, we obtain a
lower-limit on the distance to MAXI~J1659$-$152 of $d>5.3$\,kpc, consistent
with the $d>6.1$\,kpc derived above.

Given that the Galactic latitude of MAXI~J1659$-$152 is relatively high,
$b^{\mathrm{II}} = 16.52$, this would place the black-hole approximately at
$z=1.8$\,kpc, which is above average, but not unprecedented (for example
XTE~J1118$+$480 is at $z=1.6$\,kpc, \citealt{Jonker04}) and would place
MAXI~J1659$-$152 in a sub-class of high-$z$ short period galactic BHB, along with
XTE~J1118$+$480, GRO~J0422$+$32 (e.g. \citealt{Filippenko95}) and
Swift~J1753.6$-$0127 (e.g. \citealt{Zurita08}) in the Galactic Halo.

\cite{Kong10} suggest a counterpart for MAXI~J1659$-$152 with $r \simeq 22.4$, based
on archival data taken before the outburst. Given a color correction of
$V-R = 1.8$ for an M5 star, and the expected extinction of $A_R = 1.621$
\citep{schlegel1998:ApJ500}, we find an estimated $V$ magnitude for this
counterpart of $V=22.5$ corrected for extinction. If we assume
$d>6.1$\,kpc, we derive an absolute magnitude for the companion of $M_V <
8.6$, too bright for an M5 star.

\cite{Shahbaz98} find a relationship between the magnitude of the change in
optical brightness of an X-ray transient in outburst and the orbital
period, which predicts $\Delta V = 11.5$ for a $2.4$\,hour orbital
period. The maximum UVOT observed $v$-band brightness of MAXI~J1659$-$152
was $v=16.5$, which would predict a quiescent optical brightness of
$v=28$. Correcting for extinction and assuming a distance of $d>6.1$\, kpc,
gives an absolute magnitude estimate for the quiescent optical emission of
$M_V < 12.2$, consistent with the estimate that the companion star is an M5
dwarf, but inconsistent with the suggested $r$-band counterpart. We
therefore believe that the counterpart proposed by \cite{Kong10} is likely
an unrelated foreground object.

The rise to peak and slow decay of the optical emission appears to be
anti-correlated with the measurement of the spectral hardness of the
source, suggesting a strong link between the optical brightness and the
emission state of MAXI~J1659$-$152. This, along with the similar outburst
profile of the hard X-ray BAT and UVOT light-curves, suggests that the
majority of the optical emission is correlated to the Comptonized component
of the spectrum (e.g. from an optically thin jet that becomes fainter as
the X-ray spectrum becomes more thermally dominated, \citealt{Markoff01}),
rather than thermal optical emission from the disk itself. The correlated
X-ray/optical behavior seen in MAXI~J1659$-$152 is similar to the behavior observed
previously in GX~339$-$4 \citep{Homan05}, although no rapid drop in the
optical emission was detected as in that BHB. As the drop in optical flux
in GX~339$-$4 was consistent with the state transition to the Thermal
State, the lack of such a drop in MAXI~J1659$-$152 is consistent with the fact that
no such state transition was observed, and that the source remained in an
intermediate state for the remainder of the \swift\ observations.

MAXI~J1659$-$152 shows a rapid rise of the measured absorption in the initial part
of the outburst, which we believe is the evolution of a localized absorbing
medium, rather than a statistical artifact of the fitting.  This observed
increase of intrinsic absorption in the early stage outburst may be an
indication of absorption by a thermally driven wind from the accretion disk
(e.g. \citealt{ProgaandKallman2002}) or possibly a magnetic driven wind
(e.g. \citealt{Miller06}).  An evolving disk wind, has also been suggested
as a mechanism for variable absorption seen in AGN \citep{Gierlinski04}.
However, we note that high resolution X-ray spectral studies of the BHB
have suggested that disk winds are not present in BHBs in the Hard State
(e.g. \citealt{Blum10}), casting doubt on the disk-wind hypothesis during
the early part of the outburst.

Variable absorption has been seen during the initial part of the outburst
of the transient GS~2023$+$338 \citep{Zycki99}, who suggested a powerful
outflowing wind was the cause of the intrinsic emission.  Unfortunately
\swift/XRT data are not of sufficiently high spectral energy resolution to
resolve the expected absorption features of a wind. If the source of the
emission is due to ionized plasma above the disk, which has been suggested
to be a common feature of LMXBs from observations of dipping sources
\citep{DiazTrigo06}, it may also be the case that such an intrinsic
absorption change may only be seen in sources with relatively high
inclinations.

The fact that this absorption increase is seen only in the very earliest
part of the rapid rise to outburst, means that it is likely a poorly
studied phenomenon, given the difficulty in catching this early part of the
outburst with an X-ray telescope with sufficient low-energy sensitivity to
measure absorption, and the relatively short-lived timescale of the change
($<2$\,days). Therefore we believe that even if this is a common feature of
BHB outbursts, it would not be common to observe this feature without the
rapid triggering and slewing capabilities of an soft X-ray observatory such
as \swift, even if the rise of intrinsic absorption is a common feature in
BHBs that are entering outburst. In this case, the \swift\ dataset on
MAXI~J1659$-$152 represents a unique high cadence monitoring of the initial stages
of a newly discovered BHB in outburst. For comparison another BHB that
triggered BAT, Swift~J1753.5$-$0127, was only observed once for a short
1.3\,ks observation with XRT \citep{Burrows05b} in the first 6 days after
it was discovered by BAT \citep{Palmer05}.

The disk component of the spectrum shows significant evolution during the
outburst, both in the fitted temperature and normalization. The
normalization of the {\tt diskbb} model is equivalent to $\kappa =
(R_\mathrm{in}/d_{10}^2) cos(i)$, where $R_\mathrm{in}$ is in units of
kilometers, and $d_{10}$ is in units of 10 kpc, therefore although we
cannot directly measure the inner radius without making assumptions about
the inclination and distance to MAXI~J1659$-$152, changes in the
normalization during the outburst give a direct measure of the magnitude of
the change of the inner disk radius as a function of both time and
temperature.

Between MJD 55465 and MJD 55475 the value of $kT_\mathrm{in}$, the inner disk
temperature, rises from $\sim0.25$ to $\sim0.8$\,keV. In that time the
normalization reduces from $\sim 3 \times 10^4$ on MJD 55466 to $\sim 320$,
suggesting that the inner disk radius reduces by a factor of $\sim 10$ in
approximately 3\,days.  Studies of the inner disk radius of BHBs during in
different emission states have suggested a correlation of the fitted
$R_\mathrm{in}$ and $kT_\mathrm{in}$, and the suggested dependence of radius and
temperature is given by $kT_\mathrm{in} \propto R_\mathrm{in}^{-3/4}$
\citep{Cabanac09}.

\begin{figure}
\resizebox{\hsize}{!}{\includegraphics[angle=270]{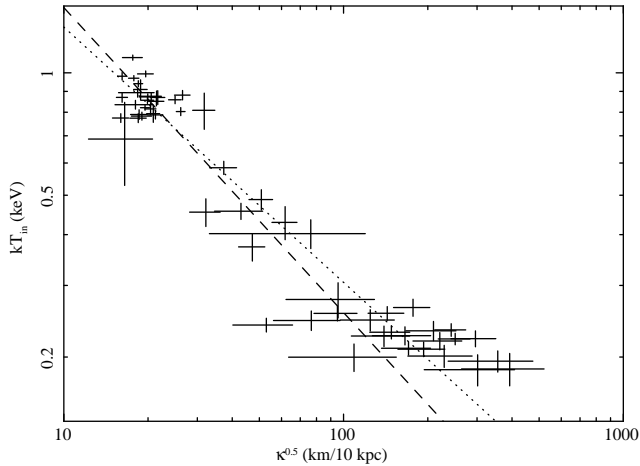} }
\caption{Evolution of the fitted inner disk temperature ($kT_\mathrm{in}$) as a
  function of the square root of the model normalization, which is
  proportional to the inner disk radius of the accretion disk. The best fit
  lines are plotted, the dotted line is the best fit power-law model, with
  a slope of $-0.62 \pm 0.02$. The dashed line is the fitted model with
  the power-law slope fixed at a value of $-3/4$, the theoretical expected
  value as proposed by \cite{Cabanac09}.}
\label{fig:rootnormvstemp}
\end{figure}

In Figure~\ref{fig:rootnormvstemp}, we plot $\kappa^{0.5}$ versus the
$kT_\mathrm{in}$, to demonstrate this correlation between the fitted
temperature and the inner disk radius for MAXI~J1659$-$152.  The best fit power-law
model to these data is $-0.62\pm 0.02$, lower than the expected $-0.75$,
although the data are broadly consistent with the expected trend in the
$kT_\mathrm{in} = 0.4-1$\,keV range. We clearly observe a trend of a decrease in
the inner disk radius both with rising $kT_\mathrm{in}$ and with
time, suggesting that accretion disk is undergoing a rapid in-fall during
the outburst of MAXI~J1659$-$152.

In the period where the flux from the accretion disk is the dominant
spectral component (MJD 55485 - MJD 55491, the latter date being due to the
end of \swift\ observations), the average {\tt diskbb} normalization is
$\kappa = 525 \pm 22$, and is statistically consistent with being constant
during this time.  If we assume that this minimum normalization corresponds
an accretion disk with an $R_\mathrm{in}$ at the innermost stable circular
orbit (ISCO) of a non-spinning black-hole, then $R_\mathrm{in}/R_\mathrm{g}
= 6$, where $R_\mathrm{g} = GM_1/c^2$.  We can estimate the mass of the
black-hole by re-arranging Equation 4 of \cite{Cabanac09} as $M_1 = 0.677\
(R_\mathrm{in}/R_\mathrm{g})^{-1}\ d_{10}\ (\kappa/cos(i))^{0.5}\
\mathrm{M_\odot}$. Utilizing this equation, the assumption that the inner
disk radius is at the ISCO, our previously obtained lower-limit on the
distance to MAXI~J1659$-$152 of $d > 6.1$\,kpc, and the allowed range of
inclinations of $i=60-75^\circ$, we find an estimated black-hole mass of
$M_1 = 2.2\ \mathrm{M_\odot} (i=60^\circ)$ to $3.1\ \mathrm{M_\odot}
(i=75^\circ)$.  This estimated mass is strongly dependent on the assumed
radius of the ISCO, which in the case of a maximally spinning black-hole,
can be as small as $1 R_\mathrm{g}$, therefore if we assume
$R_\mathrm{in}/R_\mathrm{g} = 1$, this increases the estimated BH mass
to $M_1 = 13.4 (i=60^\circ)$ to $18.6 \mathrm{M_\odot} (i=75^\circ)$.

Larger distances increase the estimate of the black-hole mass. However,
given the high galactic latitude, larger distances become increasingly
unlikely due to the derived large distance of MAXI~J1659$-$152 above the Galactic
Plane, e.g.\ $z>2.0$\,kpc for $d>7$\,kpc.

\section{Conclusion}

\swift\ and \maxi\ have discovered a new BHB, MAXI~J1659$-$152. Broadband
high-cadence \swift\ monitoring of the outburst revealed the key spectral
and timing signatures of BHBs, including the presence of time variable QPOs
and spectral evolution of the transient through canonical states associated
with BHBs. MAXI~J1659$-$152 has similarities to other BHBs, for example
Swift~J1753.5$-$0127, which has a similar outburst light-curve
\citep{Soleri10} and the presence of a low-temperature disc component, with
$kT_\mathrm{in} \simeq 0.2$ keV, during the Hard State
\citep{Miller06b}.  The increasing flux contribution of the thermal disk
component and rise of $kT_\mathrm{in}$ to $0.8-1$\,keV suggests
evolution to the Thermal State from the Steep Power-Law State, although no
full state transition was seen before observations ended.  As the disk
component becomes more prominent and hotter, the radius of the inner disk
edge of the accretion disk shrinks, approximately following the expected
$kT_\mathrm{in} \propto R_\mathrm{in}^{-3/4}$ dependence predicted by
\cite{Cabanac09}.

During the first day of observations, we recorded a rapid increase in the fitted
$N_\mathrm{H}$, likely caused by localized absorption from an evolving disk
wind. Optically, MAXI~J1659$-$152 shows a correlated rise with X-ray during
the Hard State with brightness peaking at the transition between the Hard
and Steep Power-Law states, followed by a decrease in the optical
brightness while in an intermediate state.

We detect a $\sim2.4$\,hour periodicity in the X-ray light-curve. The
structure of the associated dips is highly variable from orbit to orbit,
suggesting periodic obscuration from the accretion disk or stream. If we
assume this is the orbital period, it would make MAXI~J1659$-$152 the
shortest orbital period BHB yet known, the previous being
Swift~J1753.5$-$0127 at 3.2\,hours \citep{Zurita08}. Given the likely
distance of $d>6.1$\,kpc, and the high galactic latitude, it is likely that
MAXI~J1659$-$152, similar to Swift~J1753.5$-$0127 and other short period BHBs, is a
Galactic Halo BHB.

\acknowledgements

This work is support by NASA grant NNX10AK40G, through the \swift\ Guest
Investigator Program.  PR and VM acknowledge financial contribution from
the agreement ASI-INAF I/009/10/0. APB and PAE acknowledge the support of the
UK Space Agency. This work made use of data supplied by the UK Swift Science Data
Centre at the University of Leicester. We acknowledge the use of public
data from the \swift\ data archive.


\begin{thebibliography}{}
\bibitem[Barthelmy et al.(2005)]{Barthelmy05} Barthelmy, S.~D., et 
al.\ 2005, \ssr, 120, 143 

\bibitem[Belloni et al.(2010)]{Belloni10} 
Belloni, T.~M., Motta, S., Mu{\~n}oz-Darias, T.\ 2010, The Astronomer's Telegram, 2926

\bibitem[Blum et al.(2010)]{Blum10} Blum, J.~L., Miller, 
J.~M., Cackett, E., Yamaoka, K., Takahashi, H., Raymond, J., Reynolds, 
C.~S., \& Fabian, A.~C.\ 2010, \apj, 713, 1244 


\bibitem[Burrows et al.(2005a)]{Burrows05} Burrows, D.~N., et al.\ 
2005, \ssr, 120, 165 

\bibitem[Burrows et al.(2005b)]{Burrows05b} Burrows, D.~N., 
Racusin, J., Morris, D.~C., Roming, P., Chester, M., Verghetta, R.~L., 
Markwardt, C.~B., 
\& Barthelmy, S.~D.\ 2005, The Astronomer's Telegram, 547

\bibitem[Cabanac et al.(2009)]{Cabanac09} Cabanac, C., Fender, 
R.~P., Dunn, R.~J.~H., K\"ording, E.~G.\ 2009, \mnras, 396, 1415 

\bibitem[Chen et al.(1997)]{Chen97} Chen, W., Shrader, C.~R., 
\& Livio, M.\ 1997, \apj, 491, 312 

\bibitem[de Ugarte Postigo et al.(2010)]{deUP10} de Ugarte 
Postigo, A., Flores, H., Wiersema, K., Thoene, C.~C., Fynbo, J.~P.~U., 
\& Goldoni, P.\ 2010, GRB Coordinates Network, Circular Service, 11307 

\bibitem[D{\'{\i}}az Trigo et al.(2006)]{DiazTrigo06} D{\'{\i}}az
  Trigo, M., Parmar, A.~N., Boirin, L., M{\'e}ndez, M., \& Kaastra, J.~S.\
  2006, \aap, 445, 179


\bibitem[Eachus et al.(1976)]{Eachus76} Eachus, L.~J., Wright, 
E.~L., \& Liller, W.\ 1976, \apjl, 203, L17 

\bibitem[Evans et al.(2009)]{Evans09} Evans, P.~A., et al.\ 
2009, \mnras, 397, 1177 

\bibitem[Filippenko et al.(1995)]{Filippenko95} Filippenko, A.~V., 
Matheson, T., \& Ho, L.~C.\ 1995, \apj, 455, 614 

\bibitem[Frank et al.(1987)]{Frank87} Frank, J., King, A.~R., \& Lasota,
  J.-P.\ 1987, \aap, 178, 137

\bibitem[Frank et al.(2002)]{Frank02} Frank, J., King, A., \& Raine, D.~J.\
  2002, Accretion Power in Astrophysics (3rd Edition), Cambridge, UK: Cambridge
  University Press

\bibitem[Gehrels et al.(2004)]{Gehrels04} Gehrels, N., et al.\ 
2004, \apj, 611, 1005

\bibitem[Gierli{\'n}ski 
\& Done(2004)]{Gierlinski04} Gierli{\'n}ski, M., \& Done, C.\ 2004, \mnras, 349, L7 

\bibitem[Goad et 
al.(2007)]{Goad07} Goad, M.~R., et al.\ 2007, \aap, 476, 1401 

\bibitem[Homan et al.(2005)]{Homan05} Homan, J., Buxton, M., 
Markoff, S., Bailyn, C.~D., Nespoli, E., 
\& Belloni, T.\ 2005, \apj, 624, 295 

\bibitem[Jonker \& Nelemans(2004)]{Jonker04} Jonker, P.~G., \&
  Nelemans, G.\ 2004, \mnras, 354, 355

\bibitem[Kalamkar et al.(2011)]{Kalamkar11} Kalamkar, M., Homan, 
J., Altamirano, D., van der Klis, M., Casella, P., 
\& Linares, M.\ 2011, \apjl, 731, L2 

\bibitem[Kalberla et al.(2005)]{Kalberla05} 
Kalberla, P.~M.~W., Burton, W.~B., Hartmann, D., Arnal, E.~M., Bajaja, 
E., Morras, R., P\"oppel, W.~G.~L.\ 2005, \aap, 440, 775 

\bibitem[Kalemci et al.(2005)]{Kalemci05} Kalemci, E., Tomsick, 
J.~A., Buxton, M.~M., Rothschild, R.~E., Pottschmidt, K., Corbel, S., 
Brocksopp, C., \& Kaaret, P.\ 2005, \apj, 622, 508 

\bibitem[Kennea et al.(2010)]{Kennea10} Kennea, J.~A., et al.\ 2010, The
  Astronomer's Telegram, 2877

\bibitem[King et al.(1996)]{King96} King, A.~R., Kolb, U., 
\& Burderi, L.\ 1996, \apjl, 464, L127 

\bibitem[Kong et al.(2010)]{Kong10} Kong, A.~K.~H., et al.\ 
2010, The Astronomer's Telegram, 2976

\bibitem[Krimm et al.(2006)]{Krimm06} Krimm, H., et al.\ 2006, 
The Astronomer's Telegram, 904

\bibitem[Kuulkers et al.(2010a)]{Kuulkers10a} Kuulkers, E.,
Kouveliotou, C., van der Horst, A.~J.\ 2010a, The Astronomer's
Telegram, 2887

\bibitem[Kuulkers et al.(2010b)]{Kuulkers10b} Kuulkers, E., et al.\ 2010b, The Astronomer's
Telegram, 2912

\bibitem[Leahy et al.(1983)]{Leahy83} Leahy, D.~A., Darbro, W., 
Elsner, R.~F., Weisskopf, M.~C., Kahn, S., Sutherland, P.~G., 
\& Grindlay, J.~E.\ 1983, \apj, 266, 160 

\bibitem[Lomb(1976)]{Lomb76} Lomb, N.~R.\ 1976, \apss, 39, 447 

\bibitem[Mangano et al.(2010)]{Mangano10} Mangano, V., Hoversten, 
E.~A., Markwardt, C.~B., Sbarufatti, B., Starling, R.~L.~C., 
\& Ukwatta, T.~N.\ 2010, GRB Coordinates Network, Circular Service, 11296

\bibitem[Makishima et al.(1986)]{Makishima86} Makishima, K., 
Maejima, Y., Mitsuda, K., Bradt, H.~V., Remillard, R.~A., Tuohy, I.~R., 
Hoshi, R., \& Nakagawa, M.\ 1986, \apj, 308, 635 

\bibitem[Markoff et al.(2001)]{Markoff01} Markoff, S., Falcke, H., \&
  Fender, R.\ 2001, \aap, 372, L25

\bibitem[Matsuoka et al.(2009)]{Matsuoka09} Matsuoka, M., et al.\ 
2009, \pasj, 61, 999 

\bibitem[Miller et al.(2006a)]{Miller06} Miller, J.~M., Raymond, 
J., Fabian, A., Steeghs, D., Homan, J., Reynolds, C., van der Klis, M., 
\& Wijnands, R.\ 2006a, \nat, 441, 953 

\bibitem[Miller et al.(2006b)]{Miller06b} Miller, J.~M., Homan, J., \& Miniutti, G.\ 2006b, \apjl, 652, L113 

\bibitem[Miller et al.(2010)]{Miller10} Miller, J.~M. et al.\ 2010,
  \apj, 724, 1441 

\bibitem[Monet et al.(2003)]{Monet03} Monet, D.~G., et al.\ 
2003, \aj, 125, 984 

\bibitem[Negoro(2009)]{Negoro09} Negoro, H.\ 2009, Astrophysics 
with All-Sky X-Ray Observations, 60 

\bibitem[Negoro et al.(2010)]{Negoro10} Negoro H., et al.\ 2010, The
  Astronomer's Telegram, 2873

\bibitem[Palmer et al.(2005)]{Palmer05} Palmer, D.~M., 
Barthelmey, S.~D., Cummings, J.~R., Gehrels, N., Krimm, H.~A., Markwardt, 
C.~B., Sakamoto, T., 
\& Tueller, J.\ 2005, The Astronomer's Telegram, 546

\bibitem[Paragi et al.(2010)]{Paragi10} 
Paragi, Z., et al.\ 2010, The Astronomer's Telegram, 2906

\bibitem[Patterson(1984)]{Patterson84} Patterson, J.\ 1984, \apjs, 
54, 443 

\bibitem[Pei(1992)]{Pei92} Pei, Y.~C.\ 1992, \apj, 395, 130 

\bibitem[\protect\citeauthoryear{{Poole} et~al.}{{Poole}
 et~al.}{2008}]{poole2008:MNRAS383}
{Poole}, T.~S., et~al. 2008, \mnras, 383, 627

\bibitem[Proga \& Kallman(2002)]{ProgaandKallman2002} 
Proga, D., \& Kallman, T.~R.\ 2002, \apj, 565, 455 

\bibitem[Roming et al.(2005)]{Roming05} Roming, P.~W.~A., et 
al.\ 2005, \ssr, 120, 95  

\bibitem[Remillard 
\& McClintock(2006)]{McandRem06} Remillard, R.~A., \& McClintock, J.~E.\ 2006, \araa, 44, 49 

\bibitem[Scargle(1982)]{Scargle82} Scargle, J.~D.\ 1982, \apj, 
263, 835 

\bibitem[\protect\citeauthoryear{{Schlegel}, {Finkbeiner}, \&
 {Davis}}{{Schlegel} et~al.}{1998}]{schlegel1998:ApJ500}
{Schlegel}, D.~J., {Finkbeiner}, D.~P.,  \& {Davis}, M. 1998, \apj, 500, 525

\bibitem[Shahbaz 
\& Kuulkers(1998)]{Shahbaz98} Shahbaz, T., \& Kuulkers, E.\ 1998, \mnras, 295, L1 

\bibitem[Smith \& Dhillon(1998)]{Smith98} Smith, D.~A., \& Dhillon, V.~S.\
  1998, \mnras, 301, 767

\bibitem[Soleri et al.(2010)]{Soleri10} Soleri, P., et al.\ 
2010, \mnras, 406, 1471 

\bibitem[van der Horst et al.(2010)]{vdH10} van der Horst, A.~J., et
  al.\ 2010, The Astronomer's Telegram, 2874

\bibitem[van der Klis(1989)]{vanderKlis89} van der Klis, M.\ 1989, 
Timing Neutron Stars, 27 

\bibitem[van Paradijs 
\& McClintock(1994)]{vanParadijs94} van Paradijs, J., \& McClintock, J.~E.\ 1994, \aap, 290, 133 

\bibitem[Vovk et al.(2010)]{vovk10} Vovk, I., et al.\ 2010, The
  Astronomer's Telegram, 2875

\bibitem[Wilms et al.(2000)]{Wilms00} Wilms, J., Allen, A., 
\& McCray, R.\ 2000, \apj, 542, 914 

\bibitem[Zurita et al.(2008)]{Zurita08} Zurita, C., Durant, M., 
Torres, M.~A.~P., Shahbaz, T., Casares, J., 
\& Steeghs, D.\ 2008, \apj, 681, 1458 

\bibitem[{\.Z}ycki et al.(1999)]{Zycki99} {\.Z}ycki, P.~T., 
Done, C., \& Smith, D.~A.\ 1999, \mnras, 309, 561 

\end{thebibliography}
\end{document}